\documentclass[conference]{IEEEtran}
\IEEEoverridecommandlockouts \makeatletter
\def\ps@headings{%
	\def\@oddhead{\mbox{}\scriptsize\rightmark \hfil \thepage}%
	\def\@evenhead{\scriptsize\thepage \hfil \leftmark\mbox{}}%
	\def\@oddfoot{}%
	\def\@evenfoot{}}
\makeatother
\usepackage{graphicx,subfigure}
\usepackage{tikz}
\usetikzlibrary{topaths,calc}
\usepackage{amsmath,amsthm,amsfonts,amssymb}
\usepackage{cite}
\usepackage{bm}
\usepackage{bbm}
\usepackage{url}
\usepackage{array}
\usepackage{color,soul}
\usepackage{multirow}
\usepackage{booktabs,multirow}
\usepackage{enumerate}
\usepackage{enumitem}
\usepackage{makecell}
\usepackage{boondox-cal}

\usepackage[english]{babel}
\usepackage{algorithm}
\usepackage[noend]{algpseudocode}

\theoremstyle{plain}
\newtheorem{thm}{Theorem}[section]

\newtheorem{lem}[thm]{Lemma}

\newtheorem{defi}{Definition}[section]

\newenvironment{NewProof}{{\noindent\it Proof.}}{\hfill $\blacksquare$\par}
\newcommand{\RR}{I\!\!R}



\makeatletter
\newcommand{\ssymbol}[1]{^{\@fnsymbol{#1}}}
\newcommand{\algmargin}{\the\ALG@thistlm}
\makeatother

\usepackage{mathtools}

\allowdisplaybreaks[4]

\begin{document}
\title{A Hypergraph Approach to Distributed Broadcast}

\author{Qi Cao, Yulin Shao, Fan Yang, Octavia A. Dobre
\thanks{Q. Cao is with Xidian-Guangzhou Research Institute, Xidian University, Guangzhou, China (email: caoqi@xidian.edu.cn).}
\thanks{Y. Shao and F. Yang are with the Department of Electrical and Electronic Engineering, University of Hong Kong, Hong Kong S.A.R. (E-mail: ylshao@hku.hk).}
\thanks{O. A. Dobre is with the Faculty of Engineering and Applied Science, Memorial University, St. John’s, NL A1B 3X5, Canada (e-mail:
odobre@mun.ca).}
}

\maketitle
\pagestyle{empty}
\thispagestyle{empty}

\begin{abstract}
This paper explores the distributed broadcast problem within the context of network communications, a critical challenge in decentralized information dissemination. We put forth a novel hypergraph-based approach to address this issue, focusing on minimizing the number of broadcasts to ensure comprehensive data sharing among all network users. The key contributions of this work include the establishment of a general lower bound for the problem using the min-cut capacity of hypergraphs, and a distributed broadcast for quasi-trees (DBQT) algorithm tailored for the unique structure of quasi-trees, which is proven to be optimal. This paper advances both network communication strategies and hypergraph theory, with implications for a wide range of real-world applications, from vehicular and sensor networks to distributed storage systems.
\end{abstract}


\section{Introduction}
In the dynamically advancing field of network communications, efficiently distributing information across various nodes without centralized oversight presents a significant challenge \cite{Decentralized1,FL,Decentralized2}. As networks grow in complexity and size, the demand for cutting-edge solutions capable of managing the high demands of information dissemination both efficiently and reliably becomes increasingly crucial \cite{Decentralized2,EIC,shao2024theory}.

This paper explores the critical issue of distributed broadcast, a scenario in which each network user holds a segment of the total data and must broadcast this information to their peers. The primary challenge is determining the minimal number of broadcasts necessary to ensure that all participants acquire the complete dataset, thus achieving comprehensive network-wide information sharing. The importance of solving the distributed broadcast problem is underscored by its applications in diverse fields such as vehicular ad hoc networks \cite{tonguz2010dv,V2X}, large-scale sensor networks \cite{sensor,Age},  distributed storage, and coded caching \cite{Decentralized2,coded_caching}. In these contexts, the ability to swiftly and reliably broadcast information to all network users in a decentralized manner is crucial. This capability not only enhances network efficiency but also plays a significant role in strengthening the resilience of communication strategies used in modern distributed systems.

The distributed broadcast problem bears similarities to the well-established index coding problem \cite{index0,index1,WBN}, which involves a single server and multiple receivers. The server must satisfy all receivers' demands via broadcast in minimal time. Unlike our distributed setting, the index coding problem is centralized, and each receiver demands only one unknown message, rather than all messages. While the index coding framework provides foundational insights, it does not directly apply to the decentralized demands of our study.

Expanding upon index coding, the authors in \cite{EIC} introduced the embedded index coding (EIC) problem, and proposed a directed graph approach. In this model, multiple nodes function both as senders and receivers, where each node aims to acquire a specific subset of messages it lacks. While this approach aligns with the distributed nature of our study, it differs in how data demands are managed. The directed graph method becomes significantly complex and less efficient as the volume of data each user requires increases. This complexity compromises performance, particularly in scenarios like our distributed broadcast problem where each user needs access to the entire dataset. Consequently, the EIC methodology is nearly inapplicable for our problem, which prompted us to develop a novel hypergraph-based approach, addressing these limitations by simplifying the complexity inherent in data distribution across all users.

Earlier attempts to tackle the distributed broadcast problem, such as those in \cite{CDE}, have established preliminary bounds on the number of necessary broadcasts and proposed algorithms for the issue. However, the results from these efforts remain rudimentary, and both the lower bounds and algorithm performances fall short when compared to the methods and findings presented in this paper.

 
The main contributions of this paper are threefold.
\begin{itemize}[leftmargin=0.45cm]
    \item We formulate the distributed broadcast problem and put forth a new hypergraph approach to solve it. Our approach not only addresses the complexities inherent in distributed broadcast but also advances hypergraph theory itself. This includes the introduction of new definitions and the derivation of hypergraph properties that facilitate efficient solutions to the problem.
    \item We establish a general lower bound for the distributed broadcast problem using the min-cut capacity of hypergraphs, providing a benchmark for evaluating the efficiency of any coding and broadcast strategy. 
    \item We focus on a specific class of hypergraphs -- the quasi-trees -- and introduce the distributed broadcast for quasi-trees (DBQT) algorithm. This algorithm is tailored to exploit the unique structure of quasi-trees, and is proven to achieve the established lower bound, confirming its optimality.
\end{itemize}

{\it Notations:} We use boldface lowercase letters to represent column vectors (e.g., $\bm{s}$), boldface uppercase letters to represent matrices (e.g., $\bm{A}$), and calligraphy letters to represent sets (e.g., $\mathcal{A}$).
The cardinality of a set $\mathcal{A}$ is denoted by $|\mathcal{A}|$.
$\RR$ is the sets of real numbers, and $\mathbb{N}^{+}$ is the set of positive integers.
$[V]\triangleq {1,2,3,...,V}$.

\section{Problem Formulation}
This section provides a rigorous formulation of the distributed broadcast problem.

\textbf{Data segments}: Assume there are $W$ segments of data, where each segment $\bm{s}_w$, $w\in[W]$, is of uniform size and represented as a vector. That is, $\bm{s}_w \in \RR^L$, where $L$ is the size of each segment and $L > W$. 
Let $\bm{W} \triangleq [\bm{s}_1, \bm{s}_2, \ldots, \bm{s}_W]$ be the matrix that contains all these segments.
The $W$ data segments representing unique messages and are independent of each other. Therefore, the columns of $\bm{W}$ are linearly independent.

\textbf{Users}: Consider $V$ users, each storing a subset of the data segments, ensuring collectively that all segments are stored across these users. For any user $v \in [V]$, let $\mathcal{A}_v$ denote the set of segments stored by user $v$.
By writing $\mathcal{A}_v$ as $\{\bm{s}_{i_1}, \bm{s}_{i_2},...,\bm{s}_{i_{\left|\mathcal{A}_v\right|}}\}$, where $i_1<i_2<...<i_{|\mathcal{A}_v|}$, we define an $L\times |\mathcal{A}_v|$ matrix $\bm{A}_v$ to represent the specific segments stored by $v$: 
$$\bm{A}_v\triangleq [\bm{s}_{i_1}, \bm{s}_{i_2},...,\bm{s}_{i_{\left|\mathcal{A}_v\right|}}].$$

\textbf{Broadcast and Collision Channels}: Time is segmented into discrete slots. During each slot, only one user can broadcast a message of length $L$ to all other users. Concurrent broadcasts by multiple users result in a collision.

The primary objective of the distributed broadcast problem is to develop a coding and broadcast strategy that ensures all data segments are transmitted to all users with the fewest possible number of broadcasts.
In this paper, we focus exclusively on linear coding schemes for the broadcast process. Specifically, each message broadcast by a user is a linear combination of the user's own data segments and the messages acquired in previous slots.

As the broadcast process progresses, each user accumulates an increasing number of messages, enabling the decoding of more data segments:
\begin{itemize}[leftmargin=0.4cm]
\item At the beginning of time slot $t$, we denote by ${\bm{A}}_v^{(t)}$ the matrix whose column vectors are the data segments already known to the $v$-th user, and $\widetilde{\bm{A}}_v^{(t)}$ the matrix whose column vectors are both the data segments and the messages received in previous slots by the $v$-th user.
\item During slot $t$, suppose that the broadcasting user is $v^{(t)}\in [V]$, and denote by $\bm{z}^{(t)}$ the vector broadcasted. Given the linear coding approach, there exists a column vector $\bm{d}^{(t)}$ such that $\bm{z}^{(t)}=\widetilde{\bm{A}}^{(t)}_{v^{(t)}} \bm{d}^{(t)}$.
\end{itemize}

To ease exposition, we further define a matrix $\widetilde{\bm{C}}_v^{(t)}$ such that $\widetilde{\bm{A}}_v^{(t)}=\bm{W}\widetilde{\bm{C}}_v^{(t)}$. When $t=0$, the initial storage $\widetilde{\bm{A}}_v^{(0)}={\bm{A}}_v^{(0)}$, hence the columns of $\widetilde{\bm{C}}_v^{(0)}$ are one-hot vectors, indicating the positions of individual data segments stored by user $v$.
As time progresses ($t>0$), we apply elementary column operations to $\widetilde{\bm{C}}_v^{(t)}$ to transform as many columns as possible into one-hot vectors. These one-hot vectors are then grouped into a submatrix denoted by $\bm{C}_v^{(t)}$. This submatrix represents the data segments that have been successfully decoded by user $v$ by the end of the $t$-th slot, thus ${\bm{A}}_v^{(t)}=\bm{W}\bm{C}_v^{(t)}$.

For any user $v\in[V]$,
$$\widetilde{\bm{A}}^{(t+1)}_v = \left[ \widetilde{\bm{A}}^{(t)}_v , \bm{z}^{(t)} \right],$$
$$\widetilde{\bm{C}}^{(t+1)}_v = \left[ \widetilde{\bm{C}}^{(t)}_v , \widetilde{\bm{C}}^{(t)}_{v^{(t)}} \bm{d}^{(t)} \right].$$    
In particular, if $\widetilde{\bm{A}}^{(t)}_v$ and $\bm{z}^{(t)}$ are linearly independent, we have ${\bm{A}}_v^{(t+1)}=\bm{W}\bm{C}_v^{(t+1)}$ by elementary column operations; otherwise, we have $\bm{C}^{(t+1)}_v = \bm{C}^{(t)}_v$ and $\bm{A}^{(t+1)}_v = \bm{A}^{(t)}_v$.
    
Based on the above framework, determining the minimum number of broadcasts, denoted as $T^*_\mathcal{A}$, involves identifying the optimal sequence of broadcasting users and their corresponding coding schemes $\{v^{(t)},\bm{z}^{(t)}\}_{t=0}^{T^*_\mathcal{A}-1}$ such that, at the conclusion of these broadcasts, all users have successfully decoded all data segments:
\begin{equation}
T^*_\mathcal{A} = \min_T\{T:\text{rank}(\widetilde{\bm{A}}^{(T)}_v)=W,\forall v\}.
\end{equation}

\section{A Hypergraph Representation}
To effectively address the complexity of the distributed broadcast problem and provide a robust analytical framework, this section introduces a hypergraph representation \cite{Hypergraph}. By defining and incorporating new definitions specific to distributed broadcasting, we can interpret our broadcasting challenge in the language of hypergraph. This interpretation allows us to explore lower bounds and sophisticated strategies and achieve deeper insights into the optimal sequencing and coding techniques required for efficient data dissemination.


\subsection{Hypergraph}
To lay the groundwork for defining the hypergraph structure, we first reformulate the system model using set-based terminology. In our current system model, we have established that $\mathcal{A}_v$ denotes the set of segments stored by user $v$. Consequently, let $\mathcal{A}\triangleq \{\mathcal{A}_1,\mathcal{A}_2,...,\mathcal{A}_V\}$ represent the storage topology, illustrating how data is distributed among users. Additionally, we define $\mathcal{W}=\{\bm{s}_1, \bm{s}_2,...,\bm{s}_W\}$ as the comprehensive set of all data segments, where $\mathcal{W}=\bigcup_{v\in [V]} \mathcal{A}_v$.

For any subset $\mathcal{S} \subseteq \mathcal{W}$, the complement is denoted by $\mathcal{S}^\mathrm{c} = \mathcal{W} \setminus \mathcal{S}$, representing the segments not included in $\mathcal{S}$.
Similarly, for any set $\mathcal{e} \subseteq \{1, 2, \ldots, V\}$, the complement is written as $\mathcal{e}^\mathrm{c} = \{1, 2, \ldots, V\} \setminus \mathcal{e}$, indicating the users not encompassed by $\mathcal{e}$.

\begin{defi}\label{defi:1}
    For any $\mathcal{e}\subseteq \{1, 2, \ldots, V\}$, we define 
    $$\mathcal{A}_\mathcal{e} \triangleq \bigcup_{v\in \mathcal{e}} \mathcal{A}_{v},$$
    $$\mathcal{S}_\mathcal{e} \triangleq \left(\bigcap_{v\in \mathcal{e}} \mathcal{A}_{v} \right) \setminus \left(\bigcup_{v\in \mathcal{e}^\mathrm{c}} \mathcal{A}_{v} \right) =\left(\bigcap_{v\in \mathcal{e}} \mathcal{A}_{v} \right) \bigcap \left(\bigcap_{v\in \mathcal{e}^\mathrm{c}} \mathcal{A}^\mathrm{c}_{v} \right),$$
where $\mathcal{S}_\mathcal{e}$ denotes the set of segments that are commonly held by the users in $\mathcal{e}$ and that are not available to any users in $\mathcal{e}^\mathrm{c}$.
\end{defi}

\begin{defi}\label{defi:2}
    Let $\mathcal{H}=(\mathcal{V,E},w)$ be a weighted hypergraph representing the initial storage topology $\mathcal{A}$ with cardinality $V$ such that
    $$\mathcal{V(H)}=\{1, 2, \ldots, V\},$$
    $$\mathcal{E(H)}=\{\mathcal{e}\subseteq \mathcal{V(H)}: \mathcal{S}_\mathcal{e}\neq \emptyset, 1< |\mathcal{e}|< V\},$$
    and the weight $w:\mathcal{E(H)}\to\mathbb{N}^{+}$. In particular, for any subset $\mathcal{E}'\subseteq \mathcal{E}$, with slight abuse of notation, we define $w(\mathcal{E}')=\sum
	_{\mathcal{e}\in \mathcal{E}'}w(\mathcal{e})$.
\end{defi}
	
Combining Definitions \ref{defi:1} and \ref{defi:2}, it becomes evident that $\forall \mathcal{e}\in \mathcal{E(H)}$, $w(\mathcal{e})=|\mathcal{S}_\mathcal{e}|$.
Let $\mathcal{A}^{(t)}\triangleq \{\mathcal{A}_1^{(t)},\mathcal{A}_2^{(t)},...,\mathcal{A}_V^{(t)}\}$ be the sets of data segments known to each user in the beginning of slot $t$. 
$\mathcal{A}^{(t)}$ can also be represented as a hypergraph $\mathcal{H}^{(t)}=(\mathcal{V,E}^{(t)},w^{(t)})$.
In this model, any edge $\mathcal{e}$ is removed from $\mathcal{E}$ if and only if the segments in $\mathcal{S}_\mathcal{e}$ become known to all users, reflecting the collective updating of segments across the network.

Given this hypergraph representation, the minimum number of broadcasts, denoted by $T^*_\mathcal{H}$, is determined by
\begin{equation}\label{eq:hypergraphP}
{\color{blue}T^*_\mathcal{H}=T^*_\mathcal{A}=\min_T\{T:\mathcal{E}^{(T)}=\emptyset\}. }
\end{equation}

\subsection{Definitions}
In addressing the challenges in \eqref{eq:hypergraphP}, we now introduce several new definitions specifically tailored to our problem to facilitate the identification of optimal user selection and coding strategies. 
Examples are given later in Section \ref{sec:exp}.

\subsubsection{Partial Hypergraph \& Induced Subhypergraph}\label{sec:hypergraph}
	
A hypergraph $(\mathcal{V',E'},w)$ is called a \emph{partial hypergraph} of $\mathcal{H}=(\mathcal{V,E},w)$ if $\mathcal{V'}\subseteq \mathcal{V}$ and $\mathcal{E'}\subseteq \mathcal{E}$.
Moreover, if $\mathcal{E}'=\{\mathcal{e}:\mathcal{e}\in \mathcal{E},\mathcal{e}\subseteq\mathcal{V'}\}$, $\mathcal{H}_\mathcal{V'}\triangleq(\mathcal{V',E'},w)$ is called the \emph{largest partial hypergraph} of $\mathcal{H}$ dictated by $\mathcal{V'}$. For any partial hypergraph $(\mathcal{V',E''},w)$ of $\mathcal{H}$, we have $\mathcal{E''}\subseteq \mathcal{E'}$.
	
A hypergraph $(\mathcal{V',E'},w')$ is called \emph{induced subhypergraph} of $\mathcal{H}=(\mathcal{V,E},w)$ if 
	\begin{itemize}
		\item $\mathcal{V'}\subseteq \mathcal{V}$;
		\item $\mathcal{E}'=\{\mathcal{e}\cap \mathcal{V'}:\mathcal{e}\in \mathcal{E} \text{ and } |\mathcal{e}\cap \mathcal{V'} |\ge 2\}$;
		\item $\forall \mathcal{e'}\in \mathcal{E'}$, 
		$w'(\mathcal{e'})=w\left(\{\mathcal{e}\in \mathcal{E}:\mathcal{e}\cap \mathcal{V'}=\mathcal{e'} \}\right).$
	\end{itemize}
We will also say that $\widetilde{\mathcal{H}}_\mathcal{V'}\triangleq(\mathcal{V',E'},w')$ is the subhypergraph of $\mathcal{H}$ induced by $\mathcal{V'}$.

\subsubsection{Degree \& weighted degree}

Given $\mathcal{H}=(\mathcal{V,E},w)$, $\forall v\in \mathcal{V}$, let $\mathcal{H}[v]$ denote the set of edges connecting $v$:
$$\mathcal{H}[v]\triangleq \{\mathcal{e}:v\in \mathcal{e},\mathcal{e}\in \mathcal{E} \}.$$
The \emph{degree} of $v$ is defined as $d_\mathcal{H}(v)\triangleq |\mathcal{H}[v]|$ and the \emph{weighted degree} of $v$ is defined as $\widetilde{d}_\mathcal{H}(v)\triangleq w(\mathcal{H}[v])$.

\subsubsection{Path \& Loose Path}
An alternating sequence
$$(v_1,\mathcal{e}_1,v_2,\mathcal{e}_2,\dots,v_n,\mathcal{e}_n,v_{n+1})$$ of vertices $v_1,v_2,\dots,v_n$ and edges $\mathcal{e}_1,\mathcal{e}_2,\dots,\mathcal{e}_n$, satisfying that $v_i,v_{i+1}\in \mathcal{e}_i\in \mathcal{E}$ for $1\le i\le n$, is called a \emph{walk} connecting $v_1$ and $v_{n+1}$, or, a $(v_1,v_{n+1})$-walk.
A walk is called a \emph{path} if all edges and vertices are distinct, in which case we call it a $(v_1,v_{n+1})$-path.
A path is a \emph{cycle} if and only if $v_1=v_{n+1}$.
A path is a \emph{loose path} if $\mathcal{e}_i\cap \mathcal{e}_{j+1}=\emptyset$ for $1\le i\le n, \mathcal{e}_i\cap \mathcal{e}_{i+1}={v_i}$, and $1\le i<j\le n-1$.

\subsubsection{Connected, Tree, Quasi-tree}
A hypergraph is \emph{connected} if for any two distinct vertices, there is a walk connecting these two vertices.
A connected hypergraph with no cycles is called a \emph{tree}. 

\begin{defi}\label{defi:tree}
    Given a connected hypergraph $\mathcal{H}=(\mathcal{V,E},w)$, if any partial hypergraph $(\mathcal{V,E'},w)$ of $\mathcal{H}$ is not connected, where $\mathcal{E'}\subset \mathcal{E}$,
then $\mathcal{H}$ is called a quasi-tree.
\end{defi}

A tree is a quasi-tree, yet a quasi-tree is not necessarily a tree. For any two distinct vertices in a tree, there must be a loose path connecting them.     

\subsubsection{Cut}
Given $\mathcal{H}=(\mathcal{V,E},w)$, let $\mathcal{X}_1,\mathcal{X}_2,...\mathcal{X}_I$, $I\in\mathbb{N}$ and $I\ge 2$, be a sequence of nonempty subsets of $\mathcal{V}$. Denote the set of edges connecting these subsets by
$$\mathcal{H}[\mathcal{X}_1,\mathcal{X}_2,...,\mathcal{X}_I] \triangleq\{\mathcal{e}\in \mathcal{E(H)}:\mathcal{e}\cap \mathcal{X}_i\neq \emptyset, \forall i\in[I]\}$$

A \emph{cut} of $\mathcal{H}$ is defined as $\dot{\mathcal{H}}\mathcal{[X]} \triangleq \mathcal{H[X,V\setminus X]}$,
where $\mathcal{X}$ is nonempty and $\mathcal{X}\subset \mathcal{V}$. The weight of the cut is defined as $\delta_\mathcal{H} (\mathcal{X}) \triangleq w(\dot{\mathcal{H}}\mathcal{[X]}).$
A \emph{min-cut} of a hypergraph $\mathcal{H}$ is a cut with the minimum weight. The \emph{min-cut capacity} of $\mathcal{H}$ is the weight of a min-cut of $\mathcal{H}$, and is denoted by
$$\Delta_\mathcal{H}\triangleq \min_{\mathcal{X}\subset \mathcal{V(H)} \atop \mathcal{X}\neq \emptyset } \delta_\mathcal{H} (\mathcal{X}).$$

 	\begin{figure}[t]
		\centering 
		\begin{tikzpicture}
			\node (v1) at (0,2) {};
			\node (v2) at (1.5,3) {};
			\node (v3) at (4,2.5) {};
			\node (v4) at (0,0) {};
			\node (v5) at (2,0.5) {};
			\node (v6) at (3.5,0) {};

			\begin{scope}[fill opacity=0.8]
				\filldraw[fill=yellow!70] ($(v1)+(-0.5,0)$) 
				to[out=90,in=180] ($(v2) + (0,0.5)$) 
				to[out=0,in=90] ($(v3) + (1,0)$)
				to[out=270,in=0] ($(v2) + (1,-0.8)$)
				to[out=180,in=270] ($(v1)+(-0.5,0)$);
				\filldraw[fill=blue!60] ($(v4) + (-0.5,-0.2)$) 
				to[out=90,in=270] ($(v1)+(-0.3,-1)$)
				to[out=90,in=180] ($(v1)+(0,0.5)$)
				to[out=0,in=90] ($(v1)+(0.7,0)$)
				to[out=270,in=90] ($(v4)+(0.4,-0.2)$)
				to[out=270,in=270] ($(v4)+(-0.5,-0.2)$);
				\filldraw[fill=green!80] ($(v5)+(-0.5,0)$)
				to[out=90,in=225] ($(v3)+(-0.5,-1)$)
				to[out=45,in=270] ($(v3)+(-0.7,0)$)
				to[out=90,in=180] ($(v3)+(0,0.5)$)
				to[out=0,in=90] ($(v3)+(0.7,0)$)
				to[out=270,in=90] ($(v3)+(-0.3,-1.8)$)
				to[out=270,in=90] ($(v6)+(0.5,-0.3)$)
				to[out=270,in=270] ($(v5)+(-0.5,0)$);
				\filldraw[fill=red!70] ($(v2)+(-0.5,-0.2)$) 
				to[out=90,in=180] ($(v2) + (0.2,0.4)$) 
				to[out=0,in=180] ($(v3) + (0,0.3)$)
				to[out=0,in=90] ($(v3) + (0.3,-0.1)$)
				to[out=270,in=0] ($(v3) + (0,-0.3)$)
				to[out=180,in=0] ($(v3) + (-1.3,0)$)
				to[out=180,in=270] ($(v2)+(-0.5,-0.2)$);		
				\filldraw[fill=gray!70] ($(v4)+(0.8,-0.2)$) 
				to[out=225,in=270] ($(v4) + (-0.4,-0.2)$) 
				to[out=90,in=180] ($(v4) + (0.,0.4)$) 
				to[out=0,in=180] ($(v5) + (0,0.3)$)
				to[out=0,in=90] ($(v5) + (0.5,-0.2)$)
				to[out=270,in=0] ($(v5) + (0,-0.4)$)
				to[out=180,in=45] ($(v4)+(.8,-0.2)$);
			\end{scope}

			\fill (v1) circle (0.1) node [right] {$v_1$};
			\fill (v2) circle (0.1) node [below left] {$v_2$};
			\fill (v3) circle (0.1) node [left] {$v_3$};
			\fill (v4) circle (0.1) node [below] {$v_4$};
			\fill (v5) circle (0.1) node [below right] {$v_5$};
			\fill (v6) circle (0.1) node [below left] {$v_6$};
		\end{tikzpicture}
		\caption{\centering{An example of a hypergraph $\mathcal{H}=(\mathcal{V,E},w)$ .
  }}\label{fig.1}
	\end{figure}

\subsection{Examples}\label{sec:exp}

Fig.~\ref{fig.1} gives an example of a hypergraph $\mathcal{H}=(\mathcal{V,E},w)$, where $\mathcal{V}=\{v_1,v_2,v_3,v_4,v_5,v_6\}$ , $\mathcal{E}=\{\{v_1,v_2,v_3\},\allowbreak\{v_2,v_3\},\allowbreak\{v_1,v_4\},\allowbreak\{v_4,v_5\},\allowbreak\{v_3,v_5,v_6\}\}$, and the weights of edges are all 1.

If $\mathcal{V'}=\{v_1,v_2,v_3\}$, the largest partial hypergraph of $\mathcal{H}$ dictated by $\mathcal{V'}$ is $\mathcal{H}_\mathcal{V'}=(\mathcal{V',E'},w)$, where $\mathcal{E'}=\{\{v_1,v_2,v_3\},\{v_2,v_3\}\}$.
If $\mathcal{V''}=\{v_2,v_3,v_6\}$, the subhypergraph of $\mathcal{H}$ induced by $\mathcal{V''}$ is $\widetilde{\mathcal{H}}_\mathcal{V''}=(\mathcal{V'',E''},w'')$, where $\mathcal{E''}=\{\{v_2,v_3\},\{v_3,v_6\}\}$, $w''(\{v_2,v_3\})=2$, and $w''(\{v_3,v_6\})=1$.

For user $v_1$, the set of edges connecting $v_1$ is $\mathcal{H}[v_1]\triangleq \{\{v_1,v_2,v_3\},\{v_1,v_4\}\}.$ The degree of $v_1$ is $d_\mathcal{H}(v_1)\triangleq 2$ and the weighted degree of $v_1$ is $\widetilde{d}_\mathcal{H}(v)\triangleq 2.$

The hypergraph $\mathcal{H}$ in Fig.~\ref{fig.1} is connected, but it is not a tree because there is a $(v_2,v_3)$-cycle.
For a connected hypergraph, we can generate the partial hypergraphs by removing one or more edges. For example, by removing the edge $\{v_1,v_2,v_3\}$ in $\mathcal{H}$, we can get a partial hypergraph of $\mathcal{H}$ denoted by $\mathcal{H}'$, as shown in Fig.~\ref{fig.2}. This hypergraph is still connected, so $\mathcal{H}$ is not a quasi-tree.
    
For the connected hypergraph $\mathcal{H}'$, the partial hypergraph obtained by removing any edge in $\mathcal{H}'$  is no longer connected. Thus, $\mathcal{H}'$ is a quasi-tree. Furthermore, $\mathcal{H}'$ is also a spanning quasi-tree of $\mathcal{H}$ .

Moveover, in the hypergraph $\mathcal{H}$, let $\mathcal{X}=\{v_4,v_5,v_6\}$. Then, a cut of $\mathcal{H}$ is $\dot{\mathcal{H}}\mathcal{[X]} \triangleq \{\{v_1,v_4\},\{v_3,v_5,v_6\}\}$, the weight of which is $\delta_\mathcal{H} (\mathcal{X}) \triangleq 2.$
The min-cut of $\mathcal{H}$ is $\Delta_\mathcal{H}\triangleq 1.$

To aid the reader's understanding and provide easy reference to key concepts, Table~\ref{tab:hypergraph_defs} summarizes the main definitions and corresponding symbols used in the hypergraph, with examples based on the hypergraph in Fig.~\ref{fig.1} for clarification.

\begin{table}[t]
\centering
\caption{Hypergraph Definitions and Examples}
\label{tab:hypergraph_defs}
\footnotesize
\renewcommand{\arraystretch}{1.1}
\begin{tabular}{>{\centering\arraybackslash}m{0.15\columnwidth}>{\centering\arraybackslash}m{0.35\columnwidth}>{\centering\arraybackslash}p{0.35\columnwidth}}
\toprule
\textbf{Definitions} & \textbf{Symbols or meanings} & \textbf{Examples (from Fig.~\ref{fig.1}} \\
\midrule
Partial Hypergraph & $(V', \mathcal{E}', w)$ & $V' = \{v_1, v_2, v_3\}$, $\mathcal{E}' = \{\{v_1, v_2, v_3\}, \{v_2, v_3\}\}$ \\
\hline
Induced Subhypergraph & $\widetilde{\mathcal{H}}_{V'} = (V', \mathcal{E}', w')$ & $V' = \{v_2, v_3, v_6\}$, $\mathcal{E}'' = \{\{v_2, v_3\}, \{v_3, v_6\}\}$, \\
\hline
Degree & $d_{\mathcal{H}}(v) = |\mathcal{H}[v]|$, $\mathcal{H}[v]\triangleq \{\mathcal{e}:v\in \mathcal{e},\mathcal{e}\in \mathcal{E} \}$ & $d_{\mathcal{H}}(v_1) = 2$ \\
\hline
Weighted Degree & $\widetilde{d}_{\mathcal{H}}(v) = w(\mathcal{H}[v])$ & $\widetilde{d}_{\mathcal{H}}(v_1) = 2$ \\
\hline
Path & $(v_1, e_1, \dots, v_n, e_n, v_{n+1})$ & $(v_2,\{v_2,v_3\}, v_3,\allowbreak \{v_1,v_2,v_3\}, v_2)$ \\
\hline
Loose Path & $(v_1, e_1,\dots, v_n, e_n, v_{n+1})$ & $(v_1,\{v_1,v_4\}, v_4,\allowbreak \{v_4,v_5\}, v_5)$ \\
\hline
Tree & No cycles & Fig.~\ref{fig.1}(b) is a tree \\
\hline
Quasi-tree & No connected partial hypergraphs & Fig.~\ref{fig.1}(b) is a quasi-tree \\
\hline
Cut & $\dot{\mathcal{H}}[X] = \{e \in \mathcal{E} : e \cap X \neq \emptyset, e \cap (V \setminus X) \neq \emptyset\}$ & $X = \{v_4, v_5\}$, $\dot{\mathcal{H}}[X] = \{\{v_1, v_4\}, \{v_3, v_5, v_6\}\}$ \\
\hline
Min-cut & $\Delta_{\mathcal{H}}$: The minimum weight among all cuts & $\Delta_{\mathcal{H}} = 1$ \\
\bottomrule
\end{tabular}
\end{table}

 	\begin{figure}[t]
		\centering 
		\begin{tikzpicture}
			\node (v1) at (0,2) {};
			\node (v2) at (1.5,3) {};
			\node (v3) at (4,2.5) {};
			\node (v4) at (0,0) {};
			\node (v5) at (2,0.5) {};
			\node (v6) at (3.5,0) {};

			\begin{scope}[fill opacity=0.8]
				\filldraw[fill=blue!60] ($(v4) + (-0.5,-0.2)$) 
				to[out=90,in=270] ($(v1)+(-0.3,-1)$)
				to[out=90,in=180] ($(v1)+(0,0.5)$)
				to[out=0,in=90] ($(v1)+(0.7,0)$)
				to[out=270,in=90] ($(v4)+(0.4,-0.2)$)
				to[out=270,in=270] ($(v4)+(-0.5,-0.2)$);
				\filldraw[fill=green!80] ($(v5)+(-0.5,0)$)
				to[out=90,in=225] ($(v3)+(-0.5,-1)$)
				to[out=45,in=270] ($(v3)+(-0.7,0)$)
				to[out=90,in=180] ($(v3)+(0,0.5)$)
				to[out=0,in=90] ($(v3)+(0.7,0)$)
				to[out=270,in=90] ($(v3)+(-0.3,-1.8)$)
				to[out=270,in=90] ($(v6)+(0.5,-0.3)$)
				to[out=270,in=270] ($(v5)+(-0.5,0)$);
				\filldraw[fill=red!70] ($(v2)+(-0.5,-0.2)$) 
				to[out=90,in=180] ($(v2) + (0.2,0.4)$) 
				to[out=0,in=180] ($(v3) + (0,0.3)$)
				to[out=0,in=90] ($(v3) + (0.3,-0.1)$)
				to[out=270,in=0] ($(v3) + (0,-0.3)$)
				to[out=180,in=0] ($(v3) + (-1.3,0)$)
				to[out=180,in=270] ($(v2)+(-0.5,-0.2)$);		
				\filldraw[fill=gray!70] ($(v4)+(0.8,-0.2)$) 
				to[out=225,in=270] ($(v4) + (-0.4,-0.2)$) 
				to[out=90,in=180] ($(v4) + (0.,0.4)$) 
				to[out=0,in=180] ($(v5) + (0,0.3)$)
				to[out=0,in=90] ($(v5) + (0.5,-0.2)$)
				to[out=270,in=0] ($(v5) + (0,-0.4)$)
				to[out=180,in=45] ($(v4)+(.8,-0.2)$);
			\end{scope}

			\fill (v1) circle (0.1) node [right] {$v_1$};
			\fill (v2) circle (0.1) node [below left] {$v_2$};
			\fill (v3) circle (0.1) node [left] {$v_3$};
			\fill (v4) circle (0.1) node [below] {$v_4$};
			\fill (v5) circle (0.1) node [below right] {$v_5$};
			\fill (v6) circle (0.1) node [below left] {$v_6$};
		\end{tikzpicture}
		\caption{The partial hypergraph of $\mathcal{H}$, denoted by $\mathcal{H}'$, is a quasi-tree.
  }\label{fig.2}

\end{figure}

\subsection{A lower bound}	
Leveraging the definitions and hypergraph model established above, this section develops a lower bound for the minimum number of broadcasts.

\begin{lem}\label{lem:1}
Given a hypergraph $\mathcal{H}=(\mathcal{V,E},w)$, for any nonempty set $\mathcal{X}\subset \mathcal{V}$, we have 
\begin{equation}
    \mathcal{E}=\dot{\mathcal{H}}\mathcal{[X]}\cup \mathcal{E}(\mathcal{H}_\mathcal{X})\cup \mathcal{E}(\mathcal{H}_{\mathcal{V(H)}\setminus \mathcal{X}}).
\end{equation}
Moreover, these three sets $\dot{\mathcal{H}}\mathcal{[X]}$, $\mathcal{E}(\mathcal{H}_\mathcal{X})$ and $\mathcal{E}(\mathcal{H}_{\mathcal{V(H)}\setminus \mathcal{X}})$ are disjoint, and thus
\begin{equation}\label{eq:cutsum}
    \delta_\mathcal{H} (\mathcal{X}) + w(\mathcal{E}(\mathcal{H}_\mathcal{X} ))+w(\mathcal{E}(\mathcal{H}_{\mathcal{V(H)}\setminus \mathcal{X}}))=W.
\end{equation}
\end{lem}

\begin{thm}\label{thm:bound}  The minimum number of broadcasts $T^*_\mathcal{H}$ is bounded by
\begin{equation}
    T^*_\mathcal{H}\ge W-\Delta_\mathcal{H}.
\end{equation}
\end{thm}

\begin{NewProof}
(sketch) We first consider a disconnected hypergraph $\mathcal{H}$.
    Since $\mathcal{H}$ is disconnected, there exists a nonempty subset $\mathcal{X}\subset \mathcal{V(H)}$ such that $\dot{\mathcal{H}} [\mathcal{X}]=\emptyset$. 
    By Lemma~\ref{lem:1}, we have 
    $$w(\mathcal{E}(\mathcal{H}_\mathcal{X} ))+w(\mathcal{E}(\mathcal{H}_{\mathcal{V(H)}\setminus \mathcal{X}} ))=W.$$
    The users in $\mathcal{X}$ store $w(\mathcal{E}(\mathcal{H}_\mathcal{X} ))$ segments, and thus they need to receive $W-w(\mathcal{E}(\mathcal{H}_\mathcal{X} ))$ times at least to receive the remaining segments. Likewise, the users in $\mathcal{V(H)}\setminus \mathcal{X}$ also needs to receive $w(\mathcal{E}(\mathcal{H}_\mathcal{X} ))$ times at least. Therefore, 
    $${T}^*_\mathcal{H}\ge w(\mathcal{E}(\mathcal{H}_\mathcal{X} ))+W-w(\mathcal{E}(\mathcal{H}_\mathcal{X} ))=W.$$
    Thus, ${T}^*_\mathcal{H}=W$ if $\mathcal{H}$ is disconnected.

    Now we consider a connected hypergraph $\mathcal{H}=(\mathcal{V,E},w)$. Let $\delta_\mathcal{H}(\mathcal{X})$ be a min-cut of $\mathcal{H}$.
    Clearly $\mathcal{H'}\triangleq (\mathcal{V,E}\setminus \delta_\mathcal{H}(\mathcal{X}),w)$ is a disconnected hypergraph.
    We can further obtain $T^*_\mathcal{H'}=w(\mathcal{E})-w(\delta_\mathcal{H}(\mathcal{X}))=W-\Delta_\mathcal{H}$. Therefore,
    $T^*_\mathcal{H}\ge T^*_\mathcal{H'} = w(\mathcal{E\setminus \delta_\mathcal{H}(\mathcal{X})}) = W-\Delta_\mathcal{H}$.
\end{NewProof}


The lower bound established by Theorem~\ref{thm:bound} is demonstrably tighter than that in \cite{CDE}. While Lemma 1 in \cite{CDE} asserts that $T_\mathcal{H}^*\ge W-\min \{w(H[v]):v\in \mathcal{V}\}$, $H[v]$ is also a cut of the hypergraph $\mathcal{H}$. Thus, we have $W-\Delta_\mathcal{H}\ge W-\min \{w(H[v]):v\in \mathcal{V}\}$, indicating that our theorem provides a more restrictive lower bound.

\section{Distributed Broadcast for Quasi-tree}


The hypergraph representation equips us with a powerful analytical framework, greatly enhancing our ability to examine the complexities of the distributed broadcast problem.
In this paper, we specifically focus on a distinct class of hypergraph structures -- the quasi-trees, as defined in Definition \ref{defi:tree}.
We present the distributed broadcast for quasi-trees (DBQT) algorithm, which is meticulously crafted to complement the structural nuances of quasi-trees and is proven to be optimal.

Considering a quasi-tree $\mathcal{T}=(\mathcal{V,E},w)$, the schematic of our DBQT algorithm is summarized in Algorithm \ref{algo:DBQT}.
We first determine the sequence of broadcasting users by means of ordered representative vertices (Section \ref{sec:IVB}). Following this ordered sequence, each designated broadcaster constructs a coding matrix and transmits coded messages sequentially (Section \ref{sec:IVC}). Finally, we will show that this structured approach ensures that all necessary data segments are disseminated optimally across the network.


\begin{algorithm}[t]
    \caption{distributed broadcast for quasi-trees (DBQT)}\label{algo:DBQT}	
    \begin{algorithmic}
        \State{\textbf{Input: } A quasi-tree $\mathcal{T}=(\mathcal{V,E},w)$.}
        \State\textbf{Initialization:}
        \State	Find an ordered representative vertices $v_1^*,v_2^*,...,v_{V^*}^*$ 
        \State  Compute $\Delta_\mathcal{T}$, the weights of a min-cut of $\mathcal{T}$
        \State{$t=0$}
        \State $\mathcal{E}=\{e_1,e_2,...,e_{|\mathcal{E}|}\}$
        \State\textbf{Execution:}
        \For {$i= 1,2,\dots,{V}^*$:}
        \State $\mathcal{Z}_i=\mathcal{A}_{v_i^*}\setminus \bigcup_{j=1}^{i-1}\mathcal{A}_{v_j^*}$
        \If {$i>1$}
        \State Randomly pick an edge $\mathcal{e}_i$ in $\mathcal{T}[v^*_1,v^*_2,...,v^*_{i-1}]\cap \mathcal{T}[v^*_{i}]$
        \State Randomly pick a set $\tilde{\mathcal{S}}_{\mathcal{e}_i}\subset \mathcal{S}_{\mathcal{e}_i}$ of cardinality $\Delta_\mathcal{T}$
        \State (such a subset always exist, since $\widetilde{\mathcal{T}}_{v_1^*,v_2^*,...,v_{i}^*}$ is connected and $|\mathcal{S}_{\mathcal{e}}|\ge \Delta_\mathcal{T}$ for any $\mathcal{e}\in \mathcal{E}$ )
        \State $\mathcal{Z}_i=\mathcal{Z}_i\cup \tilde{\mathcal{S}}_{\mathcal{e}_i}$
        \EndIf
        \State $\bm{Z}_i=[\bm{s}_{i_1},\bm{s}_{i_2},...,\bm{s}_{i_{|\mathcal{Z}_i|}}]$. Here $\bm{s}_{i_1},\bm{s}_{i_2},...,\bm{s}_{i_{|\mathcal{Z}_i|}}$ are the segments in $\mathcal{Z}_i$
        \For {$\tau=1,2,...,|\mathcal{Z}_i|-\Delta_\mathcal{T}$}
        \State {$v^{(t)} = {v_i^*}$}
        \State {$\bm{z}^{(t)} = \bm{Z}_i (1^{\tau-1}, 2^{\tau-1},...,(T_i+\Delta_\mathcal{T})^{\tau-1})^\mathsf{T}$}
        \EndFor
        \EndFor
    \end{algorithmic}
\end{algorithm}

\subsection{Ordered representative vertices}\label{sec:IVB}
To start with, we first determine the optimal sequence of broadcasting users based on the concept of ordered representative vertices.


\begin{defi}\label{df:repr}
For a connected hypergraph $\mathcal{H}=(\mathcal{V,E},w)$, a vertex set $\mathcal{V}^*\subseteq \mathcal{V}$ of size $V^*$ is a \emph{representative vertex set of $\mathcal{H}$} if 
\begin{itemize}
    \item $\bigcup_{v\in \mathcal{V}^*} \mathcal{H}[v] = \mathcal{E}$,
    \item $\widetilde{\mathcal{H}}_\mathcal{V^*}$ is connected.
\end{itemize}
\end{defi}

\begin{lem}\label{lem:order}
    Let $\mathcal{V}^*$ be a representative vertex set of $\mathcal{H}$. There exists an ordered sequence of vertices $v^*_1,v^*_2,...,v^*_{V^*}$ such that $\widetilde{\mathcal{H}}_{\{v^*_1,v^*_2,...,v^*_i\}}$ is connected $\forall i\in[V^*]$.
    We call this sequence an ordered representative vertices of $\mathcal{H}$.
\end{lem}

\begin{NewProof}
    Let $\mathcal{V}_i=\{v^*_1,v^*_2,...,v^*_i\}$ for $i=1,2,...,V^*$.
    When $i=V^*$, obviously, $\widetilde{\mathcal{H}}_{\mathcal{V}^*_i}=\widetilde{\mathcal{H}}_\mathcal{V^*}$ is connected. Now we only need to prove that for any $i$, $\widetilde{\mathcal{H}}_{\mathcal{V}^*_i}$ is connected implies that there exists a ${v}^*_i$ such that $\widetilde{\mathcal{H}}_{\mathcal{V}_i\setminus \{{v}^*_i\}}$ is also connected.
    
    Let
    $v_{j_{1}},\mathcal{e}_{j_{1}},v_{j_{2}},\mathcal{e}_{j_{2}},\dots,v_{j_{(n-1)}},\mathcal{e}_{j_{(n-1)}},v_{j_{n}}$
    be a path with the longest length $n-1$ in $\widetilde{\mathcal{H}}_{\mathcal{V}_i}$, where $1 \le j_{n}\le i$ and $n\le i$.
    Now we consider $\widetilde{\mathcal{H}}_{\mathcal{V}_i\setminus \{v_{j_{1}}\}}$.
    Let $e'_j=e_j\setminus \{v_{j_{1}}\}$ for $j={j_{2}},{j_{3}},...,{j_{(n-1)}}$. We can see that $|e'_j|\ge 2$ and thus
    $v_{j_{2}},\mathcal{e}'_{j_{2}},v_{j_{3}}\dots,v_{j_{(n-1)}},\mathcal{e}'_{j_{(n-1)}},v_{j_{n}}$ is a walk in $\widetilde{\mathcal{H}}_{\mathcal{V}_i\setminus \{v_{j_{1}}\}}$, i.e., $v_{j_{2}},v_{j_{3}},...,v_{j_{n}}$ are still connected  in $\widetilde{\mathcal{H}}_{\mathcal{V}_i\setminus \{v_{j_{1}}\}}$. If any other vertex in $\mathcal{V}_i$ is connected with $v_{j_{2}}$, then by letting $v_i^*=v_{j_{1}}$ , $\widetilde{\mathcal{H}}_{\mathcal{V}_i\setminus \{v_{j_{1}}\}}$ is a connected hypergraph. So the lemma is proved. Otherwise, there exists a vertex $v_0$ not connected with $v_{j_{2}}$ in $\widetilde{\mathcal{H}}_{\mathcal{V}_i\setminus \{v_{j_{1}}\}}$. Since $v_0$ is connected with $v_{j_{2}}$ in $\widetilde{\mathcal{H}}_{\mathcal{V}_i}$ , it must be connected with $v_{j_{1}}$.
Thus, 
    $$v_0\notin \bigcup_{j={j_{1}}}^{j_{(n-1)}} \mathcal{e}_j$$
    and there exists a $(v_0,v_{j_{1}})$-path. Note we have a $(v_{j_{1}},v_{j_{n}})$-path of length $n-1$ in $\widetilde{\mathcal{H}}_{\mathcal{V}_i}$. Then we can get a $(v_0,v_{j_{n}})$-path whose length is larger than $n-1$. 
    Obviously, the path contradicts that 
    $v_{j_{1}},\mathcal{e}_{j_{1}},v_{j_{2}},\mathcal{e}_{j_{2}},\dots,v_{j_{(n-1)}},\mathcal{e}_{j_{(n-1)}},v_{j_{n}}$ is a path with the longest length in $\widetilde{\mathcal{H}}_{\mathcal{V}_i}$.
    Therefore, $\widetilde{\mathcal{H}}_{\mathcal{V}_i\setminus \{v_{j_{1}}\}}$ is a connected hypergraph.
\end{NewProof}

The procedures to find an ordered representative vertices for any connected hypergraph $\mathcal{H}=(\mathcal{V,E},w)$ are as follows:
\begin{enumerate}
\item Find a vertex $v_1$ such that for any other vertex $v'$, $\mathcal{H}[v_1]\not\subset \mathcal{H}[v']$. Then, put this vertex $v_1$ into the representative vertex set $\mathcal{V}^*$. Define a representative edge set $\mathcal{E}^*$, and let $\mathcal{E}^*=\mathcal{H}[v_1]$ and $i=2$.
\item  Find a vertex $v_i$ ,  $v_i\notin\mathcal{V}^*$ and $v_i\in\{\mathcal{v}:\mathcal{v}\in \mathcal{e},\mathcal{e}\subseteq\mathcal{E^*}\}$ such that for any other vertex $v'$, $\mathcal{H}[v_i]\not\subset \mathcal{H}[v']$ and $\mathcal{H}[v_1]\not\subset \mathcal{E}^*$. Let $\mathcal{V}^*=\mathcal{V}^*\cup {v_i}$ , $\mathcal{E}^*=\mathcal{E}^*\cup \mathcal{H}[v_i]$ and $i=i+1$.
\item Repeat step 2 until $\mathcal{E}^*=\mathcal{E}$. Let $V^*=i-1$. Then we can get a sequence of vertices $v_1,v_2,\dots,v_{V^*}$ in $\mathcal{V}^*$. 
\end{enumerate}

For any selected vertex $v_i$, $2 \le i\le V^*$, since $v_i\in\{\mathcal{v}:\mathcal{v}\in \mathcal{e},\mathcal{e}\subseteq\mathcal{E^*}\}$, it is connected with at least one vertex in $\{v_1,v_2,...,v_{i-1}\}$. Therefore, $\widetilde{\mathcal{H}}_{\{v_1,v_2,...,v_i\}}$ is connected $\forall i\in[V^*]$. The sequence we obtained is an ordered representative vertices.

As an example, consider the quasi-tree $\mathcal{H}'$ in Fig. \ref{fig.2}. Since $\mathcal{H}[v_3]$ is $\{\{v_2,v_3\},\{v_3,v_5,v_6\}\}$, which satisfies $\mathcal{H}[v_1]\not\subset \mathcal{H}[v']$ for any other vertex $v'$, we put $v_3$ into $\mathcal{V}^*$ and put $\{v_2,v_3\}$ and $\{v_3,v_5,v_6\}$ into $\mathcal{E}^*$.
Then we find $v_5$, which satisfies all the conditions in step 2. Therefore, we put $v_5$ into $\mathcal{V}^*$ and add $\{v_4,v_5\}$ into $\mathcal{E}^*$.
Similarly, we can find $v_4$, which satisfies the conditions in step 2. Therefore, we put $v_4$ into $\mathcal{V}^*$ and add $\{v_1,v_4\}$ into $\mathcal{E}^*$. At this point, $\mathcal{E}^*$ has all of the edges in $\mathcal{H}'$, hence the sequence $v_3,v_5,v_4$ is an ordered representative vertices of $\mathcal{H}'$. 
	
\subsection{Coded broadcast}\label{sec:IVC}
Given the obtained ordered representative vertices $v_1^*,\allowbreak v_2^*,\allowbreak ...,\allowbreak v_{V^*}^*$, DBQT divides the coded broadcast into $V^*$ phases.
By Lemma~\ref{lem:order}, $\widetilde{\mathcal{T}}_{\{v^*_1,v^*_2,...,v^*_i\}}$ is connected for $i=1,2,...,V^*$. Let $\mathcal{e}_i\in \mathcal{T}\left[\{v^*_1,v^*_2,...,v^*_{i-1}\},v^*_{i}\right]$ be arbitrary for $i=1,2,...,V^*$. Specially, $\mathcal{e}_1=\emptyset$ and $\mathcal{e}_i\ge \Delta_\mathcal{T}$ for $i=2,3,...,V^*$.
Let 
$$\mathcal{Z}_i = \tilde{\mathcal{S}}_{e_i}\cup \left(\mathcal{A}_{v_i^*}\setminus \cup_{j=1}^{i-1} \mathcal{A}_{v_j^*}\right)$$ be a set of segments broadcasted in Phase $i$, where $\tilde{\mathcal{S}}_{e_i}$ is an arbitrary subset of ${\mathcal{S}}_{e_i}$ with cardinality $\min\{\Delta_\mathcal{T},{\mathcal{S}}_{e_i}\}$. By writing $\mathcal{Z}_i$ as $\{\bm{s}_{j_1}, \bm{s}_{j_2},...,\bm{s}_{j_{\left|\mathcal{Z}_i\right|}}\}$, where $j_1<j_2<...<j_{|\mathcal{Z}_i|}$, we define an $L\times |\mathcal{Z}_i|$ matrix 
$$\bm{Z}_i = \left[\bm{s}_{j_1}, \bm{s}_{j_2},...,\bm{s}_{j_{\left|\mathcal{Z}_i\right|}}\right].$$

In Phase $i$, the coded messages sent by User $v_i^*$ are the columns in $\bm{Z}_i\bm{M}_{i}$ where $\bm{M}_{i}$ is a coding matrix of size $|\mathcal{Z}_i|\times (|\mathcal{Z}_i|-\Delta_\mathcal{T})$  given by
$$\bm{M}_{i}\triangleq \left[
\begin{matrix}
    1^0 & 1^1 & \cdots & 1^{|\mathcal{Z}_i|-\Delta_\mathcal{T}-1}\\
    2^0 & 2^1 & \cdots & 2^{|\mathcal{Z}_i|-\Delta_\mathcal{T}-1}\\
    \vdots  & \vdots & \ddots &\vdots \\
    |\mathcal{Z}_i|^{0} & |\mathcal{Z}_i|^{1} & \cdots &|\mathcal{Z}_i|^{|\mathcal{Z}_i|-\Delta_\mathcal{T}-1}
\end{matrix}\right].$$

\begin{lem}\label{lem:know-one-know-all}
Consider any user storing $\Delta_\mathcal{T}$ segments in $\mathcal{Z}_i$, $i=1,2,...,V^*$.
Upon receiving the columns in $\bm{Z}_i\bm{M}_{i}$, the user is able to decode all the messages in $\mathcal{Z}_i$.
\end{lem}

\begin{NewProof}
(sketch) Let $\bm{s}_{j_{k_1}},\bm{s}_{j_{k_2}},...,\bm{s}_{j_{k_{\Delta_\mathcal{T}}}}$ be the $\Delta_\mathcal{T}$ segments stored by the user, 
and $\bm{\alpha}(k)$ denote a one-hot vector of length $|\mathcal{Z}_i|$ whose $k$-th item is $1$. 
When the users receives the columns in $\bm{Z}_i\bm{M}_{i}$, it stores columns in 
$\bm{Z}_i\bm{M}'_{i}$, where $$\bm{M}'_{i}=\left[\bm{\alpha}(k_1),\bm{\alpha}(k_2),...,\bm{\alpha}(k_{\Delta_\mathcal{T}}),\bm{M}_i\right].$$
It suffices to prove that $\det (\bm{M}'_{i})\neq 0$.
Removing the first $\Delta_\mathcal{T}$ columns and $k_1,k_2,...,k_{\Delta_\mathcal{T}}$-th rows of $\bm{M}'_{i}$, we can obtain a new matrix denoted by $$\bm{M}''_{i}=\left[\begin{matrix}
        1^0& \cdots &1^{|\mathcal{Z}_i|-\Delta_\mathcal{T}-1} \\
        \vdots &\vdots & \vdots   \\
        (k_1-1)^{0}& \cdots &(k_1-1)^{|\mathcal{Z}_i|-\Delta_\mathcal{T}-1}  \\
        (k_1+1)^0& \cdots &(k_1+1)^{|\mathcal{Z}_i|-\Delta_\mathcal{T}-1} \\
        \vdots &\vdots & \vdots   \\
        (k_2-1)^{0}& \cdots &(k_2-1)^{|\mathcal{Z}_i|-\Delta_\mathcal{T}-1}  \\
        (k_2+1)^0& \cdots &(k_2+1)^{|\mathcal{Z}_i|-\Delta_\mathcal{T}-1} \\
        \vdots &\vdots & \vdots \\
        (k_{\Delta_\mathcal{T}}-1)^{0}& \cdots &(k_{\Delta_\mathcal{T}}-1)^{|\mathcal{Z}_i|-\Delta_\mathcal{T}-1}  \\
        (k_{\Delta_\mathcal{T}}+1)^0& \cdots &(k_{\Delta_\mathcal{T}}+1)^{|\mathcal{Z}_i|-\Delta_\mathcal{T}-1} \\
        \vdots &\vdots & \vdots \\
        |\mathcal{Z}_i|^{0}& \cdots &|\mathcal{Z}_i|^{|\mathcal{Z}_i|-\Delta_\mathcal{T}-1}
    \end{matrix}\right].$$
It is evident that $$|\det (\bm{M}'_{i})|=|\det (\bm{M}''_{i})|.$$
Note that $\bm{M}''_{i}$ is a Vandermonde matrix, which is full rank. Therefore, $\det (\bm{M}'_{i})\neq 0$. 
\end{NewProof}

\begin{thm}\label{thm:DBQT}
    The DBQT algorithm achieves optimality. It ensures that all $W$ data segments are known to every user after  $T_\mathcal{T}^*=W-\Delta_\mathcal{T}$ broadcasts.
\end{thm}
\begin{NewProof}
(sketch) The number of broadcasts in DBQT is 
\begin{align*}
T=&\sum_i \left(|\mathcal{Z}_i|-\Delta_\mathcal{T}\right)\\
=&|\mathcal{A}_{v_1^*}|-\Delta_\mathcal{T} +\sum_{i=2}^{V^*}|\mathcal{A}_{v_i^*}\setminus \cup_{j=1}^{i-1} \mathcal{A}_{v_j^*}|\\
=&\left|\bigcup_{i=1}^{V^*} \mathcal{A}_{v_i^*}\right|-\Delta_\mathcal{T}\\
=&W-\Delta_\mathcal{T}.
\end{align*}
By Theorem~\ref{thm:bound}, we have $T^*\ge W-\Delta_{\mathcal{T}}$.
Thus, $T\le T^*$. Now we only need to prove that each vertex $v\in \mathcal{V}$ can decode all the $W$ segments.
    We first prove that $v_1^*$ can decode any segment $\bm{s}\in\mathcal{W}$. Let 
    $J$ be the smallest such that $\bm{s}\in \bigcup_{j=1}^{J} \mathcal{A}_{v_j^*}$. (Such a $J$ always exists, since by Definition~\ref{df:repr}, $\bigcup_{j=1}^{J} \mathcal{A}_{v_j^*}=\mathcal{W}$ when $J=V^*$.) By Lemma~\ref{lem:order}, 
    $\widetilde{\mathcal{T}}_{v_1^*,v_2^*,...,v_J^*}$ is connected. Thus there exists a ($v_1^*,v_J^*$)-path
    $$v_{i_1}^{*},\mathcal{e}_{i_2},v_{i_2}^*,\dots,v_{i_{k-1}}^*,\mathcal{e}_{i_k},v_{i_k}^*$$
    in $\mathcal{T}$, where $1=i_1$, $i_k=J$ and $i_j$ is the smallest such that $e_{i_{j+1}}\in \mathcal{T}[v_{i_j}]$ for $j=k-1,k-2,...,1$. 
    Since $|\tilde{\mathcal{S}}_{\mathcal{e}_{i_2}}|\ge \Delta_\mathcal{T}$ and $\tilde{\mathcal{S}}_{\mathcal{e}_{i_2}}\subseteq \mathcal{A}_{v_1^*}\cap \mathcal{Z}_{i_2}$, by Lemma~\ref{lem:know-one-know-all}, User $v_1^*$ can decode all the messages in $\mathcal{Z}_{i_2}$, including the $\Delta_{\mathcal{T}}$ segments in $\tilde{\mathcal{S}}_{\mathcal{e}_{3}}$. Thus, it can further decode all the segments in $\mathcal{Z}_3$. Repeat this argument, user $v_1^*$ can finally decode $\bm{s}$.

    Likewise, we can also prove that any User $v$ can decode all the messages in $v_1^*$.
    Since $\mathcal{T}$ is connected, there exists a ($v,v_1^*$)-path. We can obtain that any other user $v\in \mathcal{V}$ can decode the segments stored in user $v_1^*$. Then we can further obtain that $v$ can decode all the $W$ segments.
\end{NewProof}

It is worth noting that the sequence of ordered representative vertices within DBQT is not unique. Regardless of the specific sequence of vertices chosen, the fundamental properties and performance of DBQT are maintained.


\subsection{Computational complexity of DBQT}\label{sec:IVD}
The computational complexity of the DBQT algorithm can be quantified as follows
\begin{align*}
C &= \mathcal{O}(E) + \mathcal{O}\biggl(\sum_{e=1}^E r_e\biggr) + \mathcal{O}\biggl(V^2\sum_{e=1}^E r_e\biggr) \\
&\quad + \mathcal{O}\biggl(L\sum_{i=1}^{V^*} m_i(m_i-\Delta)\biggr).
\end{align*}

where $V$ denotes the number of vertices (users), $E$ denotes the number of edges, 
$r_e$ denotes the size of the $e$-th edge (number of vertices it contains), 
$w_e$ denotes the number of segments on the $e$-th edge with $W=\sum_e w_e$, 
$\Delta_\mathcal{H}=\min_e w_e$ denotes the min-cut, 
$L$ is the length of each segment vector, 
$V^*$ denotes the number of representative vertices, 
and $m_i$ denotes the number of segments to be broadcast in the $i$-th phase. 

To be more specific:
\begin{itemize}
    \item The first term \(\mathcal{O}(E)\) is the cost of computing the min-cut, where \(\Delta_\mathcal{H}=\min_e w_e\) can be obtained by a single scan of edge weights. 
    \item The second term \(\mathcal{O}\!\big(\sum_{e} r_e\big)\) corresponds to building all \(H[v]\) sets($r_e$ denotes the number of vertices contained by edge $e$). 
    \item The third term \(\mathcal{O}\!\big(V^2\sum_e r_e\big)\) is the conservative worst-case cost of selecting the representative vertex set. In practice, since \(V^* \ll V\), this step is usually closer to \(\mathcal{O}\!\big(V\sum_e r_e\big)\). 
    \item The fourth term \(\mathcal{O}\!\big(L\sum_{i=1}^{V^*} m_i(m_i-\Delta_\mathcal{H})\big)\) is the cost of DBQT encoding and broadcasting. 
\end{itemize}

\section{Conclusions}
This paper formulated and addressed the distributed broadcast problem, a challenge with wide-reaching implications in network communications. 
We established a structured and analytical framework using a hypergraph-based representation of the storage topology.
This framework is vital for comprehending and managing the intricate interdependencies characteristic of broadcast networks. 
Our development of the DBQT algorithm marked a significant achievement, as it effectively minimized broadcast times for quasi-trees, aligning with theoretical predictions.
Our contributions lay the groundwork for both theoretical advancements and practical applications in network communications, paving the way for future innovations in distributed systems. 

\appendices

\section{DBQT on General Hypergraphs}
The optimality of the DBQT algorithm on quasi-trees has been proven in Section \ref{sec:IVC}. In this appendix, we investigate the performance of DBQT on general hypergraphs through simulations.

For a general hypergraph, we note that if the hypergraph is disconnected, the theoretical lower bound for the number of broadcasts is equal to the total number of data segments, $W$. In such cases, we can simply select a set of vertices that cover all the data segments and broadcast them directly. Therefore, for our evaluation, we focus on the non-trivial connected hypergraphs.

\begin{figure}[t]
    \centering
    \includegraphics[width=0.9\linewidth]{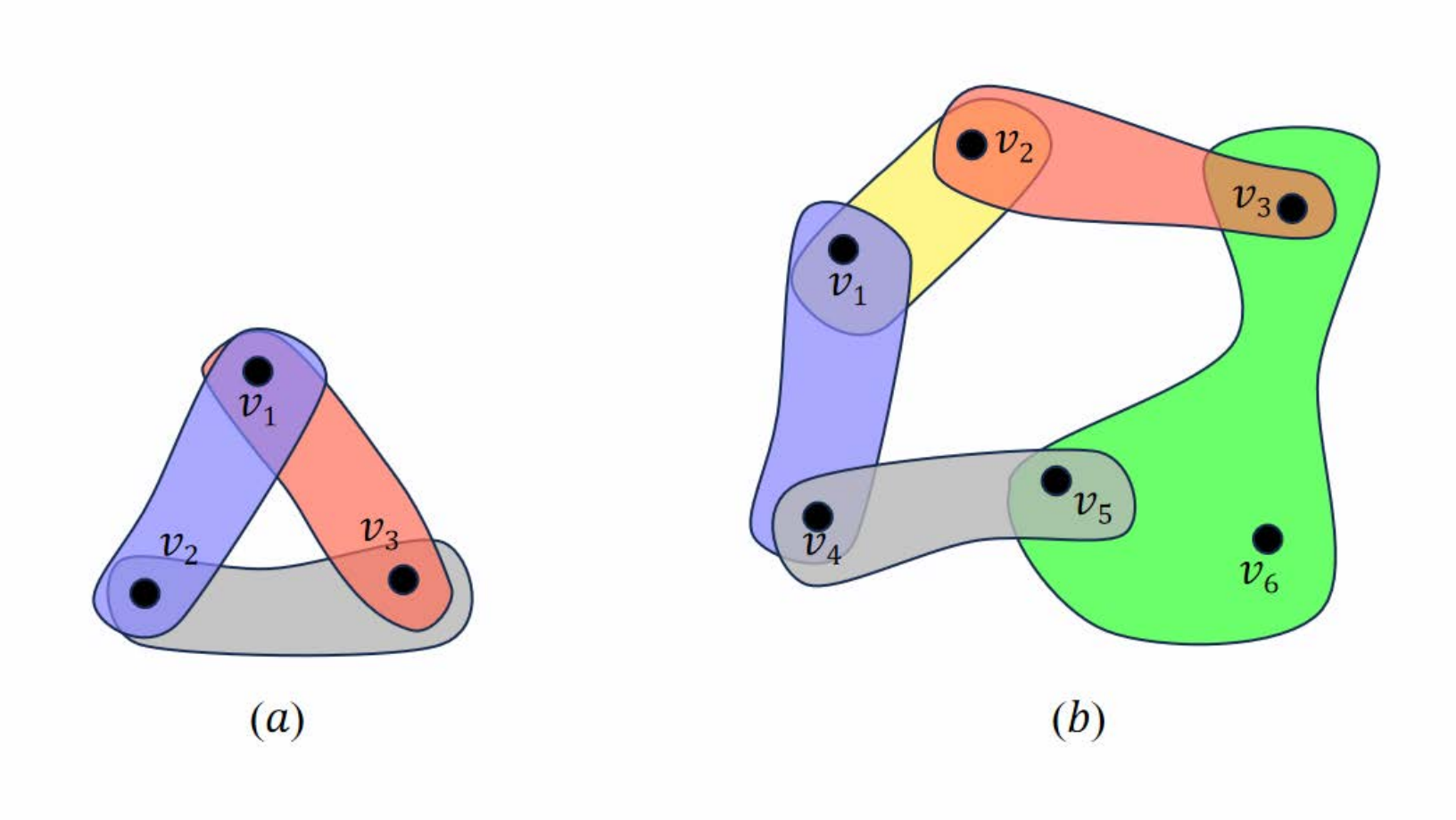}
    \caption{(a) A cycle composed of three edges.
             (b) A non-quasi-tree constructed from the quasi-tree given in Fig.~\ref{fig.2}.}
    \label{fig:general}
\end{figure}

For connected non-quasi-tree hypergraphs, we can still apply the DBQT algorithm by first generating a quasi-tree structure through the removal of certain edges. These removed edges typically form a cycle with other edges in the original hypergraph. An example of this is illustrated in Fig.~\ref{fig:general}(a). By adding cycles to the quasi-tree, we can reconstruct a general connected hypergraph. For instance, in the example from Fig.~\ref{fig.2}, the quasi-tree denoted by $\mathcal{H'}$ can be converted into a non-quasi-tree by adding an edge $\{v1,v2\}$, as shown in Fig.~\ref{fig:general}(b).

To evaluate the performance of DBQT on general connected hypergraphs, we proceed as follows: we randomly generate a quasi-tree with $V$ vertices and $W$ data segments, then add additional edges to create a non-quasi-tree. We repeat this process for $100$ randomly constructed non-quasi-trees for each pair of $V$ and $W$. The results are averaged and compared with the theoretical lower bound.

\begin{figure}[t]
  \centering
  \includegraphics[width=0.9\linewidth]{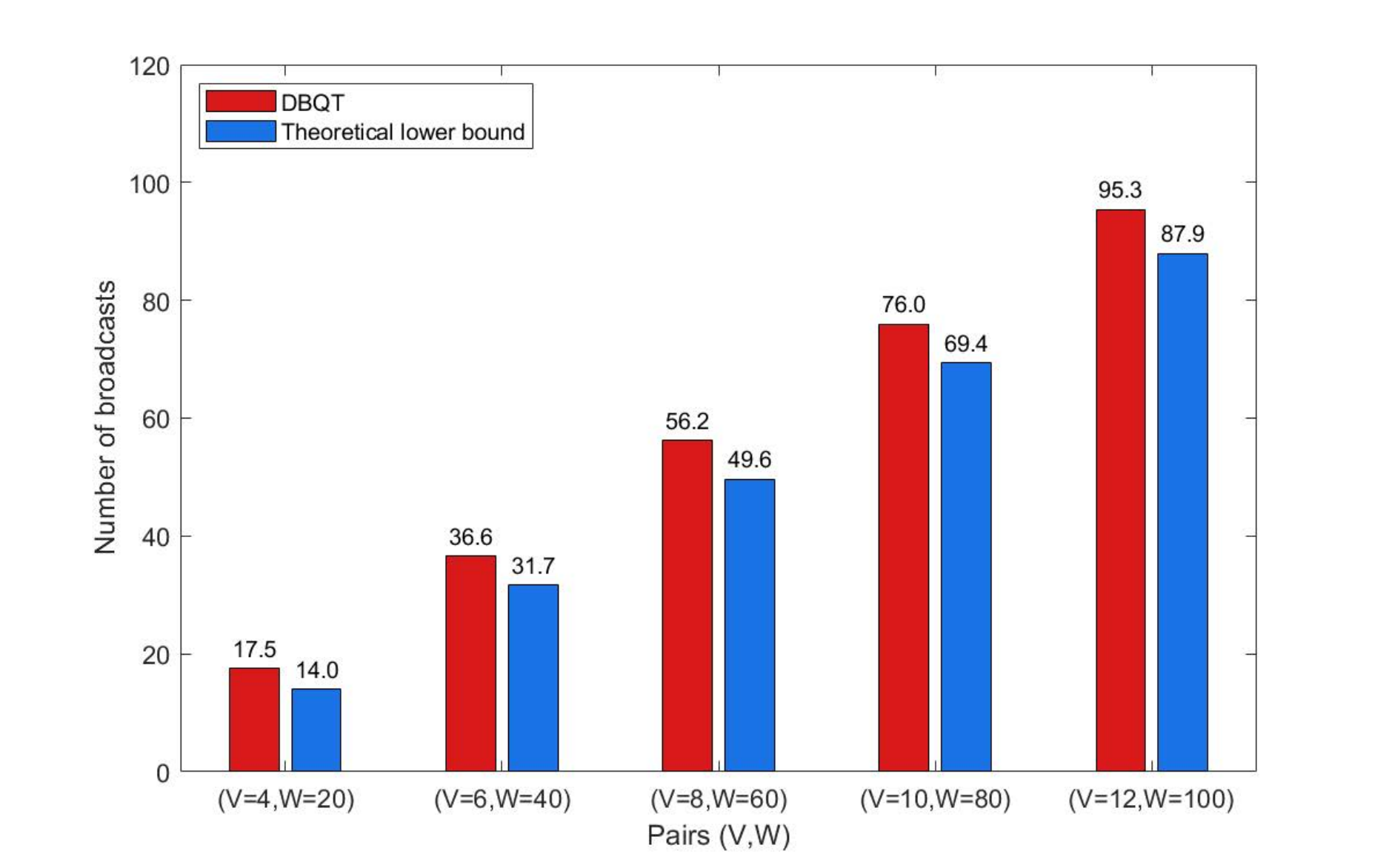}
  \caption{DBQT vs Lower bound on general hypergraphs (non-quasi-trees).}
\label{dbqt_gen}
\end{figure}

The simulation results, presented in Fig.~\ref{dbqt_gen}, demonstrate that the number of broadcasts achieved by DBQT satisfies the inequality:
\[
W- \Delta_\mathcal{H}\leq T \leq W.
\]

Importantly, although DBQT does not reach the theoretical lower bound in general hypergraphs, the gap is relatively small, suggesting that DBQT performs well even on general non-quasi-tree hypergraphs.

\bibliographystyle{IEEEtran}
\bibliography{References}
\end{document}